\documentclass[preprintnumbers, showpacs, amsmath, showkeys, amssymb, aps, superscriptaddress, twocolumn, longbibliography, letterpaper]{revtex4-2}
\usepackage{lipsum, babel}
\usepackage{color}
\usepackage{hyperref}
\usepackage{bm}
\usepackage{amsfonts}
\usepackage{mathrsfs}
\usepackage{graphicx}
\usepackage{amsmath}
\usepackage{float}
\usepackage{braket}
\usepackage{natbib}

     \hypersetup{
    colorlinks=true,
    linkcolor=red,
    citecolor=blue,
}

\begin{document}

\newcommand{\sgn}{\operatorname{sgn}}
\newcommand{\hhat}[1]{\hat {\hat{#1}}}
\newcommand{\pslash}[1]{#1\llap{\sl/}}
\newcommand{\kslash}[1]{\rlap{\sl/}#1}
\newcommand{\lab}[1]{}
\newcommand{\sto}[1]{\begin{center} \textit{#1} \end{center}}
\newcommand{\rf}[1]{{\color{blue}[\textit{#1}]}}
\newcommand{\eml}[1]{#1}
\newcommand{\el}[1]{\label{#1}}
\newcommand{\er}[1]{Eq.\eqref{#1}}
\newcommand{\df}[1]{\textbf{#1}}
\newcommand{\mdf}[1]{\pmb{#1}}
\newcommand{\ft}[1]{\footnote{#1}}
\newcommand{\n}[1]{$#1$}
\newcommand{\fals}[1]{$^\times$ #1}
\newcommand{\new}{{\color{red}$^{NEW}$ }}
\newcommand{\ci}[1]{}
\newcommand{\de}[1]{{\color{green}\underline{#1}}}
\newcommand{\ke}{\rangle}
\newcommand{\br}{\langle}
\newcommand{\lb}{\left(}
\newcommand{\rb}{\right)}
\newcommand{\lbk}{\left[}
\newcommand{\rbk}{\right]}
\newcommand{\blb}{\Big(}
\newcommand{\brb}{\Big)}
\newcommand{\nn}{\nonumber \\}
\newcommand{\p}{\partial}
\newcommand{\pd}[1]{\frac {\partial} {\partial #1}}
\newcommand{\cd}{\nabla}
\newcommand{\cc}{$>$}
\newcommand{\bqa}{\begin{eqnarray}}
\newcommand{\eqa}{\end{eqnarray}}
\newcommand{\bqe}{\begin{equation}}
\newcommand{\eqe}{\end{equation}}
\newcommand{\bay}[1]{\left(\begin{array}{#1}}
\newcommand{\eay}{\end{array}\right)}
\newcommand{\eg}{\textit{e.g.} }
\newcommand{\ie}{\textit{i.e.}, }
\newcommand{\iv}[1]{{#1}^{-1}}
\newcommand{\st}[1]{|#1\ke}
\newcommand{\at}[1]{{\Big|}_{#1}}
\newcommand{\zt}[1]{\texttt{#1}}
\newcommand{\non}{\nonumber}
\newcommand{\m}{\mu}
\def\xa{{m}}
\def\xA{{m}}
\def\xb{{\beta}}
\def\xB{{\Beta}}
\def\xd{{\delta}}
\def\xD{{\Delta}}
\def\xe{{\epsilon}}
\def\xE{{\Epsilon}}
\def\xve{{\varepsilon}}
\def\xg{{\gamma}}
\def\xG{{\Gamma}}
\def\xk{{\kappa}}
\def\xK{{\Kappa}}
\def\xl{{\lambda}}
\def\xL{{\Lambda}}
\def\xo{{\omega}}
\def\xO{{\Omega}}
\def\xvp{{\varphi}}
\def\xs{{\sigma}}
\def\xS{{\Sigma}}
\def\xt{{\theta}}
\def\xvt{{\vartheta}}
\def\xT{{\Theta}}
\def \Tr {{\rm Tr}}
\def\CA{{\cal A}}
\def\CC{{\cal C}}
\def\CD{{\cal D}}
\def\CE{{\cal E}}
\def\CF{{\cal F}}
\def\CH{{\cal H}}
\def\CJ{{\cal J}}
\def\CK{{\cal K}}
\def\CL{{\cal L}}
\def\CM{{\cal M}}
\def\CN{{\cal N}}
\def\CO{{\cal O}}
\def\CP{{\cal P}}
\def\CQ{{\cal Q}}
\def\CR{{\cal R}}
\def\CS{{\cal S}}
\def\CT{{\cal T}}
\def\CV{{\cal V}}
\def\CW{{\cal W}}
\def\CY{{\cal Y}}
\def\BC{\mathbb{C}}
\def\BR{\mathbb{R}}
\def\BZ{\mathbb{Z}}
\def\sA{\mathscr{A}}
\def\sB{\mathscr{B}}
\def\sF{\mathscr{F}}
\def\sG{\mathscr{G}}
\def\sH{\mathscr{H}}
\def\sJ{\mathscr{J}}
\def\sL{\mathscr{L}}
\def\sM{\mathscr{M}}
\def\sN{\mathscr{N}}
\def\sO{\mathscr{O}}
\def\sP{\mathscr{P}}
\def\sR{\mathscr{R}}
\def\sQ{\mathscr{Q}}
\def\sS{\mathscr{S}}
\def\sX{\mathscr{X}}

\def\slz{SL(2,Z)}
\def\slr{$SL(2,R)\times SL(2,R)$ }
\def\ads{${AdS}_5\times {S}^5$ }
\def\adst{${AdS}_3$ }
\def\sun{SU(N)}
\def\ad#1#2{{\frac \delta {\delta\sigma^{#1}} (#2)}}
\def\bqf{\bar Q_{\bar f}}
\def\nf{N_f}
\def\sunf{SU(N_f)}

\def\dcirc{{^\circ_\circ}}

\author{Morgan H. Lynch}
\email{morganlynch1984@gmail.com}
\altaffiliation{New address: Max-Planck-Institut f\"{u}r Kernphysik, Saupfercheckweg 1, 69117 Heidelberg, Germany}
\affiliation{Center for Theoretical Physics,
Seoul National University, Seoul 08826, Korea}

\title{Analysis of the CERN-NA63 radiation reaction data set, assuming the Rindler bath is composed of microscopic black holes}
\date{\today}

\begin{abstract}
In this manuscript we examine the Unruh-thermalized CERN-NA63 radiation reaction data set from the point of view of a diphoton Rindler bath. Under the assumption that these Hawking-Unruh diphoton pairs are microscopic trans-Planckian black holes, we find the resultant heat capacity describes the measured energy spectrum and is thus a dual description of the data set. Then, employing an $n$-dimensional Stefan-Boltzmann analysis, we find the power radiated by a black hole in the standard 3+1 spacetime dimensions in complete agreement with the data. Finally, we utilize this power spectrum to directly measure Newtons constant of gravitation.
\end{abstract}


\maketitle

\section{Introduction}

The predictions of particle creation by an expanding universe by Parker \cite{Parker:1968mv}, black hole evaporation by Hawking \cite{hawking1974black}, and the effervescence of particles out of an accelerated quantum vacuum by Fulling, Davies, and Unruh \cite{Davies:1976hi, Davies:1977yv, Unruh:1976db} firmly established quantum field theory in curved spacetime as an exciting avenue of research into novel gravitational phenomena. Recent efforts in particle physics have even brought the Fulling-Davies-Unruh (FDU) effect into the realm of experimental science \cite{Wistisen:2017pgr, RDKII:2016lpd, lynch2021experimental, lynch2023experimental, 10.1093/ptep/ptad157}. In particular, the high energy channeling radiation experiments carried out by CERN-NA63 \cite{Wistisen:2017pgr} have produced sufficiently large recoil accelerations that the emitted radiation has thermalized at the FDU temperature \cite{lynch2021experimental}. Moreover, the presence of thermality in the radiation is also reflected in a completely independent analysis based on Rindler horizon thermodynamics \cite{lynch2023experimental}. Given that thermality is manifestly present in the radiation as well as the horizon, we must then ask if we can find a similar signature of thermalization from the point of view of the effervescent quantum fluctuations of the Rindler bath.

In this manuscript, we formulate our analysis based on the gravitational properties of the particles which comprise the Rindler bath. We find, based on the area change of the horizon, a Rindler particle whose gravitational mass is that of a diphoton, i.e. a photon anti-photon pair. Then, under the assumption that this quantum fluctuation can be described as a microscopic or virtual trans-Planckian black hole, the resultant thermodynamics and evaporation, at the associated Hawking temperature, provide an excellent description of the thermalized NA63 data set. In particular, the heat capacity is found to be a surprisingly simple description of the energy spectrum and also provides a simple way to determine the thermalization threshold of these systems. Moreover, in analyzing the n-dimensional Stefan-Boltzmann power radiated by these trans-Planckian black holes, we find that the standard 3+1 dimensional spacetime ($n$=3 spatial dimensions), agrees with the data with a reduced chi-squared per degree of freedom below the 1 standard deviation threshold. This energy spectrum is then used to directly measure Newton's constant directly from the data set; thereby demonstrating the presence of gravitation at CERN-NA63.

\section{Rindler Horizon area change}

Black hole thermodynamics, and its application to Rindler horizons, also gives rise to an associated area change due to the flux of energy through the horizon. Although the area of the Rindler horizon is formally infinite, the area change, $dA$, is well defined and is determined by the Rindler heat flux, $dQ = d\omega_{R}$, of matter across the horizon \cite{Jacobson:1995ab, bianchi2013mechanical}. As such, we can map the second law of thermodynamics, $dQ = k_{B}T dS$, to the Rindler setting, i.e. using $T = T_{FDU}$, along with the Bekenstein-Hawking area-entropy law \cite{Bekenstein:1973ur, hawking1974black}, $S = A/(4 \ell_{p}^2)$. Note, in the Rindler setting, the heat flux and entropy are defined in terms of densities as the Rindler horizon is functional infinite in extent. We are, however, taking the ratio of these quantities so the total area factors will vanish. Then, by recalling the energy gap of an Unruh-DeWitt detector sets the Rindler frequency \cite{Cozzella:2017ckb, 2020PhRvD.101f5012P, lynch2021experimental}, we have $\omega_{R} \approx \Omega + \beta_{\perp} \omega +\frac{\omega^2}{2m}$. Here, $\Omega$ is the channeling oscillation frequency, $\beta_{\perp} \omega$ is the resonant term we directly couple to from the Rindler bath, and $\frac{\omega^2}{2m}$ is the recoil imparted on the positron. As such, we shall have the Rindler heat flux to be that of resonant Rindler frequency our accelerated charge couples to, $dQ \approx  \beta_{\perp} d\omega$. We then obtain the Rindler horizon area change,
\bqe
\frac{dA}{d(\hbar \omega)} = \frac{4 \beta_{\perp} \ell_{p}^2}{k_{B}T_{FDU}}.
\label{da}
\eqe
This area change comes directly from the second law of thermodynamics, which itself maps to the emitted photon spectrum \cite{lynch2023experimental}, $\frac{dN}{d\omega_{R}} = \alpha_{N}\frac{dS}{d\omega_{R}}=  \frac{\alpha_{N}}{4 \ell_{p}^2}\frac{dA}{d\omega_{R}}$. Here, $\alpha_{N}$, is the amount of entropy per thermal photon emitted \cite{2016PhLB..757..383A}. This area change can be integrated to yield the total amount of area generated. For a radiating positron of mass, $m$, we will have the following recoil FDU temperature \cite{lynch2021experimental, lynch2023experimental},
\bqe
T_{FDU} =\frac{(\hbar \omega)^2}{2\pi \alpha_{T} mc^2k_{B}}.
\label{fdu}
\eqe
Here, the parameter, $\alpha_{T}$, characterizes the time duration of the acceleration, $\Delta t_{a}$, in terms of the number of photon periods \cite{lynch2023experimental}, $\Delta t_{a} = \frac{\alpha_{T}}{\omega}$. Note, this formulation of the area change, Eqn. (\ref{da}), relies only on the Raychaudhuri equation and second law of thermodynamics, \cite{Jacobson:1995ab}. Using this temperature in our differential area change, Eqn. (\ref{da}), we can integrate it to determine the total area added to the horizon.  Hence,
\bqe
A_{R} = \frac{8\pi \alpha_{T} \beta_{\perp}\ell_{p}^2 mc^2}{\hbar \omega}.
\label{ar}
\eqe
Here we have a positive sign in the area change because the energy flux is going into the horizon by the positron. From this area we shall determine the mass/energy content which is associated with the area of the resultant horizon. In the case of a black hole, we can determine its mass by examining the area-temperature product. With an area given by $A = 16 \pi M^2 G^{2}/c^{4}$, and a Hawking temperature given by, $T = \frac{\hbar c^3}{8\pi G M k_{B}}$, we then find the product is given by $Ak_{B}T = 2 Mc^{2}\ell_{p}^2 $. Thus the mass associated with the black hole, or Rindler horizon, can be determined by this product. For the case of the Rindler horizon, with the recoil FDU temperature, we have the mass/energy, $E_{R} = A_{R}k_{B}T_{FDU}$, is given by,
\bqa
E_{R} &=& 2 \beta_{\perp}(\hbar \omega) \non \\
& \approx & 2  \omega_{R}.
\eqa
The fact that we have twice the photon energy, or Rindler frequency, here reflects the notion of pair creation at the horizon as being responsible for Hawking-Unruh radiation emission. This can also be viewed as the fact that detailed balance implies both the emission of particles into and absorption of particles from the Rindler bath as being the covariant description of radiation emission as measured in the laboratory \cite{Cozzella:2017ckb}. Ultimately, it seems that this diphoton pair plays a role in the subsequent dynamics. Now that we have established an area change which implies that existence of the diphoton, let us now confirm that it is indeed the source of gravitation.

\subsection{The Rindler-Einstein diphoton ring}
\begin{figure}[H]
\centering  
\includegraphics[scale=.51]{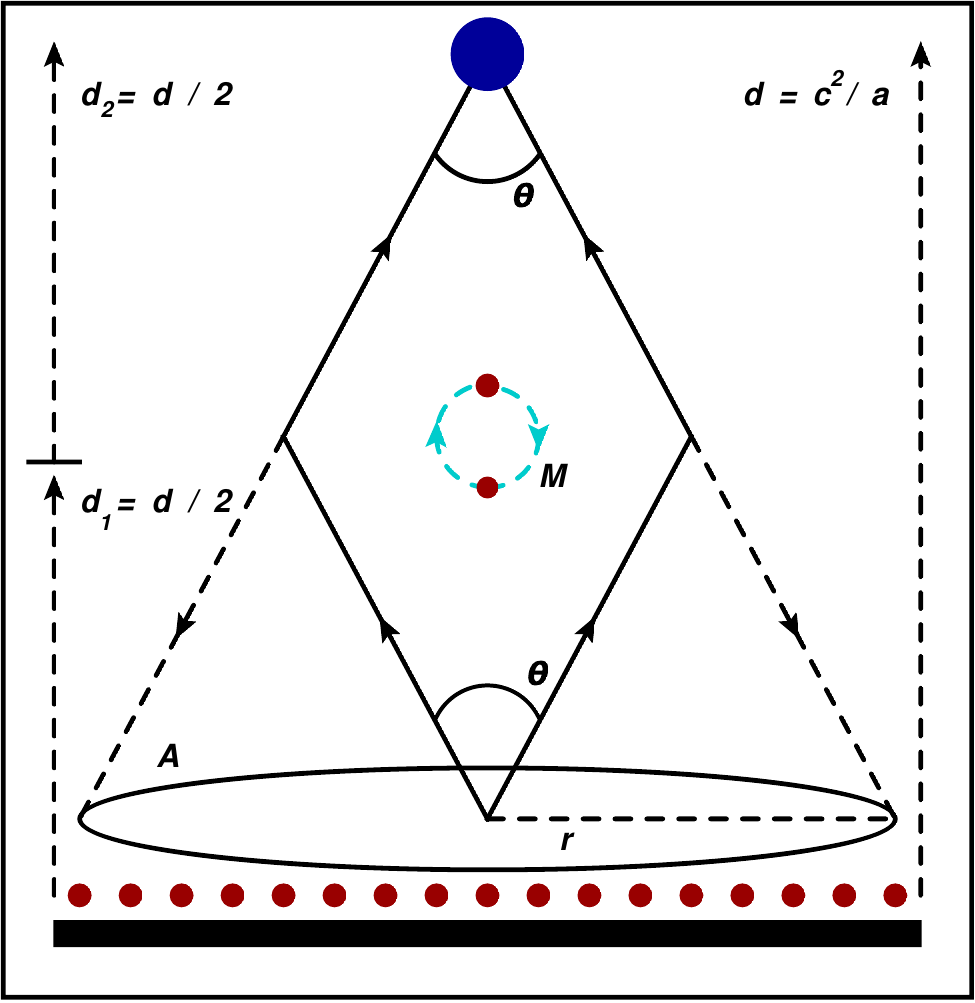}
\caption{Diagramatics of the Rindler-Einstein ring produced via the gravitational lensing by the diphoton. The area of the resultant Einstein ring is equivalent to the area generated by the flux of energy across the horizon.} 	
\label{plot1}
\end{figure}

To better understand the nature of this area change, let us look at the gravitational lensing of the system. We shall find that the area generated, or area change of the Rindler horizon, will be determined by the area of an Einstein ring which is sourced by the diphoton pair \cite{stein}. Recalling that the angular size, $\theta$, of an Einstein ring generated by a mass, $M$, is given by
\bqe
\theta = 2\lbk \frac{4G M}{c^2} \frac{d_{1}}{d_{2}d}  \rbk^{1/2}.
\eqe
In the Rindler frame, see Fig. (\ref{plot1}), the distance between the positron and the horizon is $d = c^2/a$, with $a$ being the proper acceleration of the positron. We then take the gravitating matter to be located at the midpoint between the positron and horizon. Thus the distance between the horizon and the matter, $d_{1}$, and the distance between the positron and the gravitational mass, $d_{2}$, are both given by $d_{1} = d_{2} = d/2$. Finally, the Einstein ring projected back onto the horizon will have a radius, $r = \theta d/2$, and an area given by $A = \pi \lb \frac{ \theta d}{2}\rb^{2} = 4\pi GM/a$. Upon substitution of $M  = (2 \omega_{R})/c^2 = 2 \beta_{\perp}(\hbar \omega)/c^2$ and $a = \frac{\omega^2 \hbar}{\alpha_{T}m c}$, we reproduce the area generated by the diphoton at the recoil FDU temperature, Eqn. (\ref{ar}). Note, the above formulation relied on the bending of light rays in general relativity, i.e. gravitational lensing, while the previous analysis, Eqn. (\ref{ar}), relied on geodesic deviation and the second law of thermodynamics. With the Einstein equation being an equation of state \cite{Jacobson:1995ab}, these two formulations are indeed equivalent. \textit{Both analyses indeed imply the presence of gravitation in this system in the form of the diphoton.} This pair is responsible for not only the area change, but also corresponds to that actual photons measured in the lab frame \cite{Cozzella:2017ckb, lynch2021experimental}. 

\section{Experimental methods}
The CERN-NA63 experimental site performs systematic studies of strong field QED \cite{2019PhRvR...1c3014W, 2023PhRvL.130g1601N} and here we shall focus on their high energy channeling radiation experiment which was successful in measuring radiation reaction \cite{Wistisen:2017pgr}. There, ultra-relativistic $178.2$ GeV positrons traverse a 3.8 mm thick single crystal silicon sample along the $\braket{111}$ axis. These ``channeled" positrons undergo a transverse harmonic oscillation and a photon builds up around the positron as it is excited up the ladder of harmonic oscillator states. In this manner, the photon energy becomes comparable to the positron rest mass and upon emission, the positron experiences an enormous recoil acceleration. This acceleration is sufficiently strong enough to thermalize the system via the Unruh effect \cite{lynch2021experimental,lynch2023experimental}.

In order to compare our theory to the NA63 data set, we simply transform their power spectrum, $\frac{dE}{d(\hbar \omega)dt}$ into an energy spectrum, $\frac{dE}{d(\hbar \omega)}$. With the crystal crossing time being given by, $\Delta t = (3.8$ mm)$/c$, we simply multiply the power spectrum by this time to formulate the energy spectrum. Hence, 
\bqe
\frac{dE}{d(\hbar \omega)} = \frac{dE_{data}}{d(\hbar \omega)dt} \lb \frac{3.8\;mm}{c} \rb.
\label{db}
\eqe
We must also examine the thermalization time for the system, $T_{therm}$. Thus, if we take an experimentally measured power spectrum, $\frac{d\mathcal{P}_{EXP}}{d\omega}$ \cite{Wistisen:2017pgr}, we can automatically compute its thermalization, or decay, time in a completely model independent fashion;
\bqe
T_{therm}(\omega) = \lbk \int_{0}^{\omega} \frac{d\mathcal{P}_{EXP}}{d\omega'} \frac{1}{\omega'} d\omega'\rbk^{-1} \label{time}
\eqe
A plot of this thermalization time is contained below in Fig. (\ref{plott}). Note that we indeed find a thermalization threshold at $\omega_{t}\sim$ 13 GeV. Beyond this energy scale, the system has sufficient time to thermalize.
\begin{figure}[H]
\centering  
\includegraphics[scale=.23]{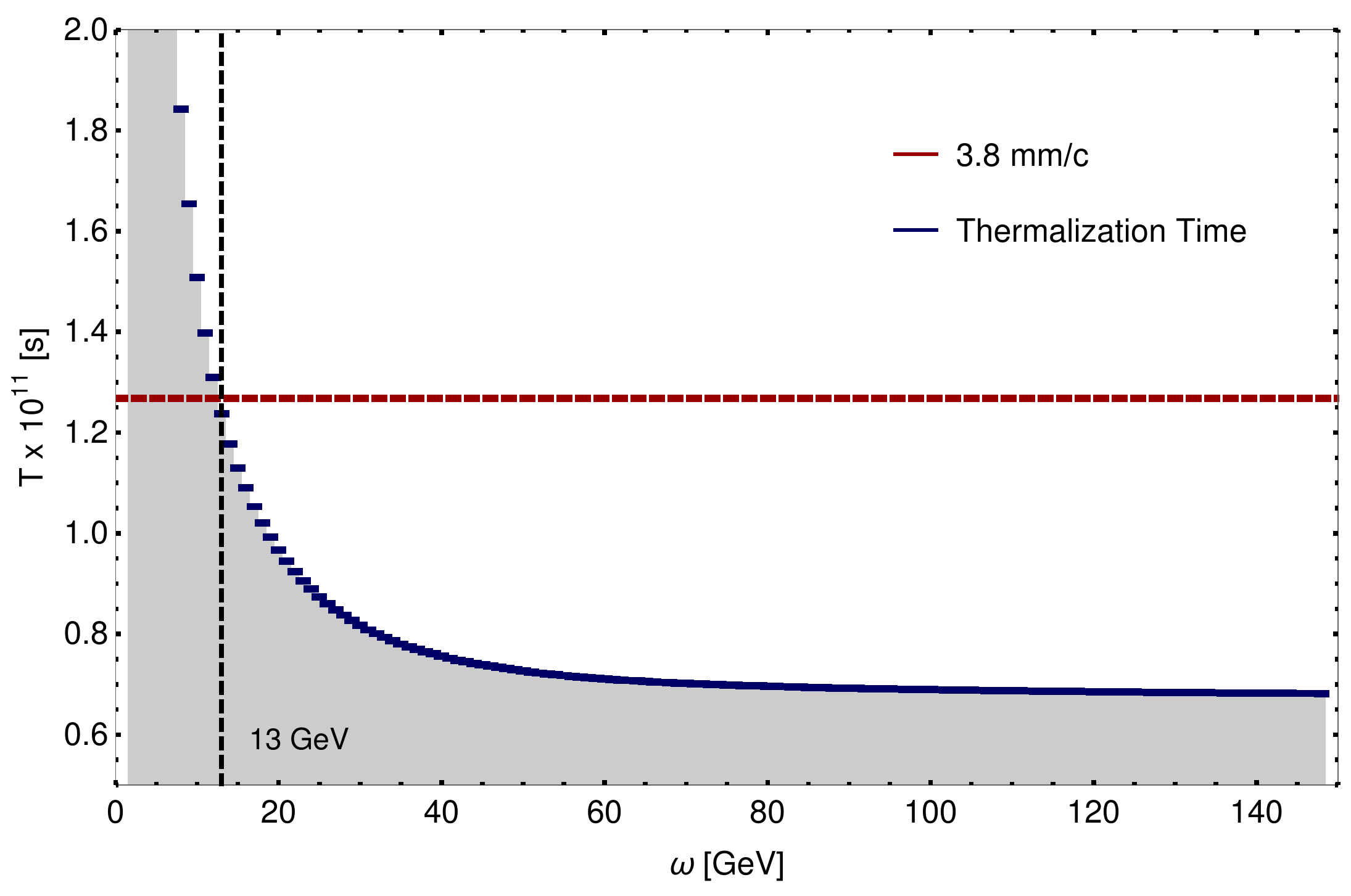}
\caption{Here we present model independent thermalization time measured the data set, Eq. (\ref{time}). We find a thermalization threshold at $\omega_{t}=$ 13 GeV.} 	
\label{plott}
\end{figure}
We will now turn to the theoretical description of this data based on aspects of Hawking radiation from the trans-Planckian diphoton black holes.

\section{N-Dimensional trans-Planckian black holes}
If we are to examine the Unruh effect from the point of view of the Rindler bath, let us note two of its key properties. First, the radiation emitted is thermal. Second, the Rindler bath appears, in this case, to be comprised of diphoton pairs which also serve as sources of gravitation. In this regard, let us consider a dual description of the Unruh effect where we assume the particles which comprise the Rindler bath are, or can be described as, trans-Planckian black holes. Thus, the creation of the photons seen in the laboratory are sourced by the decay of these diphoton black holes. With an energy of $E_{R} = 2 \omega_{R}$, we shall make use of the associated higher dimensional Myers-Perry Hawking temperature is given by \cite{1986AnPhy.172..304M},
\bqe
T_{H}=T_{P}\frac{n-2}{4\sqrt{\pi}}\lbk \frac{E_{P}}{E_{R}} \frac{n-1}{8\Gamma{\lb \frac{n}{2}\rb}}  \rbk^{\frac{1}{n-2}}
\label{temp}
\eqe
Note, here the dimension, $n$, characterizes the number of spatial dimensions and is restricted to $n \geq 3$.
This Hawking temperature is well above the Planck temperature, $T_{P} = \frac{1}{k_{B}}\sqrt{\frac{\hbar c^5}{G }} =1.4 \times 10^{32}$ K. The Rindler energy which sets the temperature is given by $\omega_{R} \approx \beta_{\perp}(\hbar \omega)$, with $\beta_{\perp} \approx .012$ \cite{lynch2021experimental, lynch2023experimental}. The photon energies measured by NA63, i.e. for energies ranging from $0-150$ GeV, will then yield a maximum $n=3$ temperature, $T_{150} = 2.0 \times10^{49}$ K.

\subsection{Heat capacity} 
\begin{figure}[H]
\centering  
\includegraphics[scale=.23]{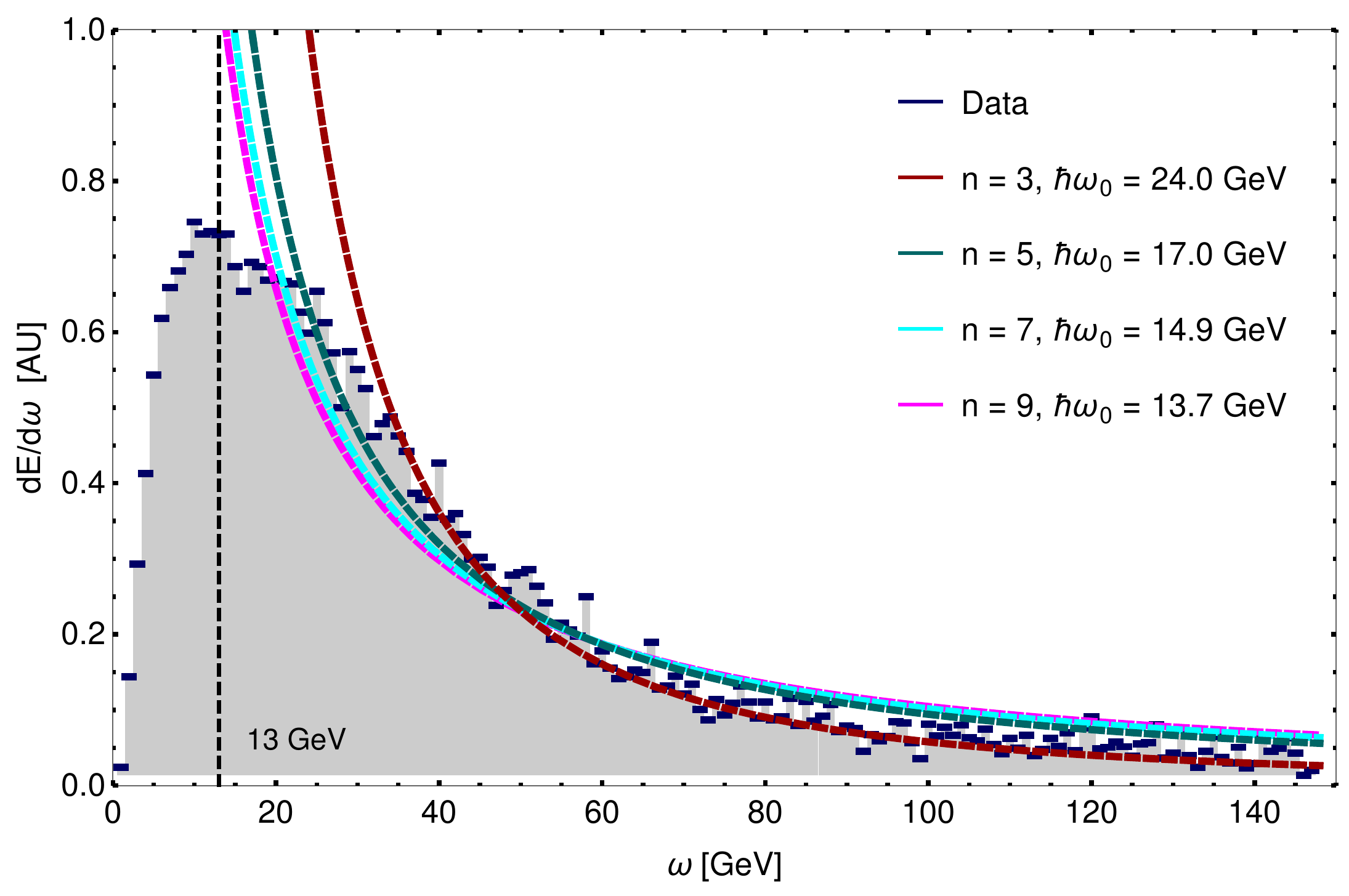}
\caption{Here we present the black hole heat capacity energy spectrum, Eqn. (\ref{cv}), for n = odd. For each of the best fit thresholds, $\hbar \omega_{0}$, we have the following chi-squared per degree of freedom $\chi^{2}_{n}/\nu$: $\chi^{2}_{3}/\nu = 1.88$, $\chi^{2}_{5}/\nu = 3.97$, $\chi^{2}_{7}/\nu = 5.91$, and $\chi^{2}_{9}/\nu = 6.90$.} 	
\label{plotodd}
\end{figure}

\begin{figure}[H]
\centering  
\includegraphics[scale=.23]{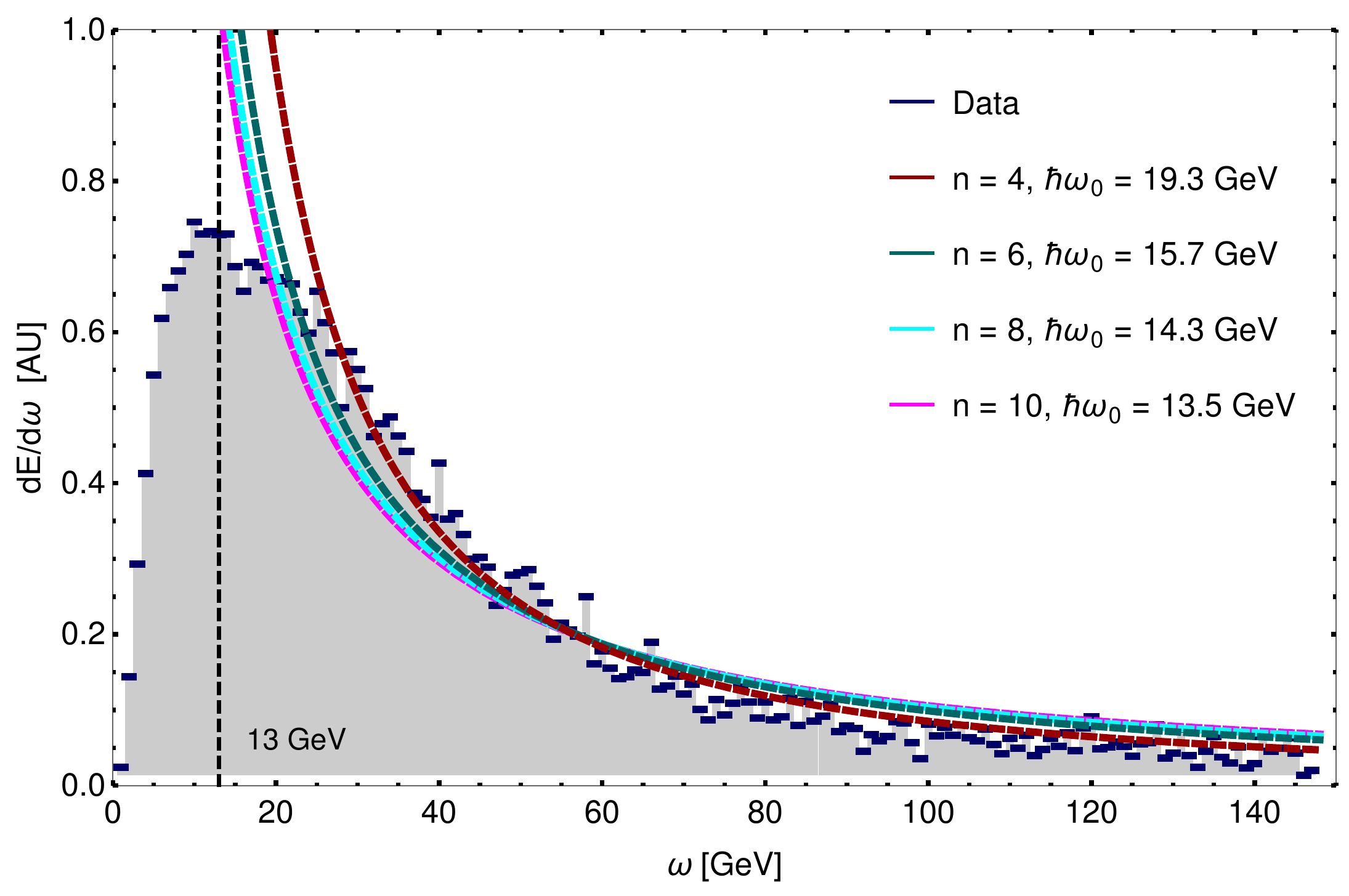}
\caption{Here we present the black hole heat capacity energy spectrum, Eqn. (\ref{cv}), for n = even. For each of the best fit thresholds, $\hbar \omega_{0}$, we have the following chi-squared per degree of freedom $\chi^{2}_{n}/\nu$: $\chi^{2}_{4}/\nu = 2.30$, $\chi^{2}_{6}/\nu = 5.12$, $\chi^{2}_{8}/\nu = 6.48$, and $\chi^{2}_{10}/\nu = 7.23$.} 	
\label{ploteven}
\end{figure}

With this dual Hawking temperature, we can examine the heat capacity at constant volume, $C_{v} = \frac{dE_{R}}{dT_{H}}$. Hence,
\bqe
C_{v} =  - \frac{k_{B}\sqrt{\pi}(n-1)}{2\Gamma{\lb \frac{n}{2} \rb}} \lbk \frac{T_{P}(n-2)}{T_{H}4\sqrt{\pi}} \rbk^{n-1}.
\eqe
Note, this relation also encodes the entropy via \cite{2022ChPhC..46e5105Z}, $C_{v} = -(n-1)S$. In order to utilize this heat capacity to examine the NA63 data set, we shall formulate a dimensionless spectrum via $\frac{d(k_{B}T_{H})}{d(\hbar \omega)} = \frac{k_{B}}{C_{v}} \frac{dE_{R}}{d(\hbar \omega)} $. This spectrum, will be ``initialized" by the ratio with the thermalization threshold $\hbar \omega_{0}$. As an example, for the standard $n=3$ case, we have $C_{v0} = -\frac{T_{P}^2k_{B}}{T_{H0}^2 8 \pi}$, and $T_{H0} = \frac{\hbar c^5}{8 \pi G (2 \beta_{\perp} \hbar \omega_{0})	 k_{B}}$. Thus, in terms of the heat capacity, we will have a relatively simple spectrum,
\bqe
\frac{dE_{C}}{d(\hbar \omega)} = \frac{C_{v0}}{C_{v}} = \lb \frac{\hbar \omega_{0}}{\hbar \omega} \rb^{\frac{n-1}{n-2}}.
\label{cv}
\eqe
This heat capacity spectrum is presented in Fig. \ref{plotodd} and \ref{ploteven} for $3 \leq n \leq 10$. We note that beyond $n >10$, the best fit threshold will be below the 13 GeV cutoff from the thermalization time, Fig. \ref{plott}. From the radiation analysis \cite{lynch2021experimental}, we have a chi-squared threshold of $\hbar \omega_{\chi} = 30$ GeV which then deviates at about $\omega = 120$ GeV. As such, our fit was performed over the data range of $30 - 120$ GeV. Below, we include the best fit threshold energies and their associated chi-squared per degree of freedom. 
Due to the fact that the signal for the standard n = 3 case is the strongest, let us also examine the relationship between the diphoton Hawking temperature, Eqn. (\ref{temp}) and the recoil FDU temperature, Eqn. (\ref{fdu}), in this setting. Specifically for the case of radiation reaction where both temperatures are functions of the photon frequency, we have
\bqe
T_{FDU} = \frac{k_{B}T_{P}^{4}}{512 \pi^3 m c^2 \beta_{\perp}^2 \alpha_{T} T_{H}^2}.
\eqe
On the Unruh side, it is the photon energy which sets the recoil acceleration and thus FDU temperature. On the Hawking side, this photon also defines the mass/energy of the diphoton black hole. Therefore, one can view the FDU temperature as a measurement of the heat capacity of the black holes which comprise the Rindler bath. Hence,
\bqe
T_{FDU} = \frac{T_{P}^{2} C_{v3}}{(8 \pi \beta_{\perp})^2 \alpha_{T} m c^2  }.
\eqe
Another intriguing aspect of this Hawking/Unruh duality is that when analyzing the power radiated by an accelerated emitter, the emission rate is determined by a thermal distribution with a Boltzmann factor, $e^{E_{R}/T_{FDU}}$. Thus the standard thermal factors which present themselves in these systems is comprised of the ratio of the Rindler energy, which characterizes the Hawking effect, and the standard FDU temperature, which characterizes the Unruh effect. In this sense, the Rindler energy/Hawking temperature defines the relevant energy of the Rindler bath microstate which we are probing in the ensemble. Again, the implication being that the Rindler bath is comprised of micrsocopic trans-Planckian black holes. Thus, by tuning the Rindler energy, i.e. $\beta_{\perp}$, one can tune the interaction so as to select the desired black hole microstate from the Rindler bath.
\subsection{N-dimensional Stefan-Boltzmann power}
\begin{figure}[H]
\centering  
\includegraphics[scale=.23]{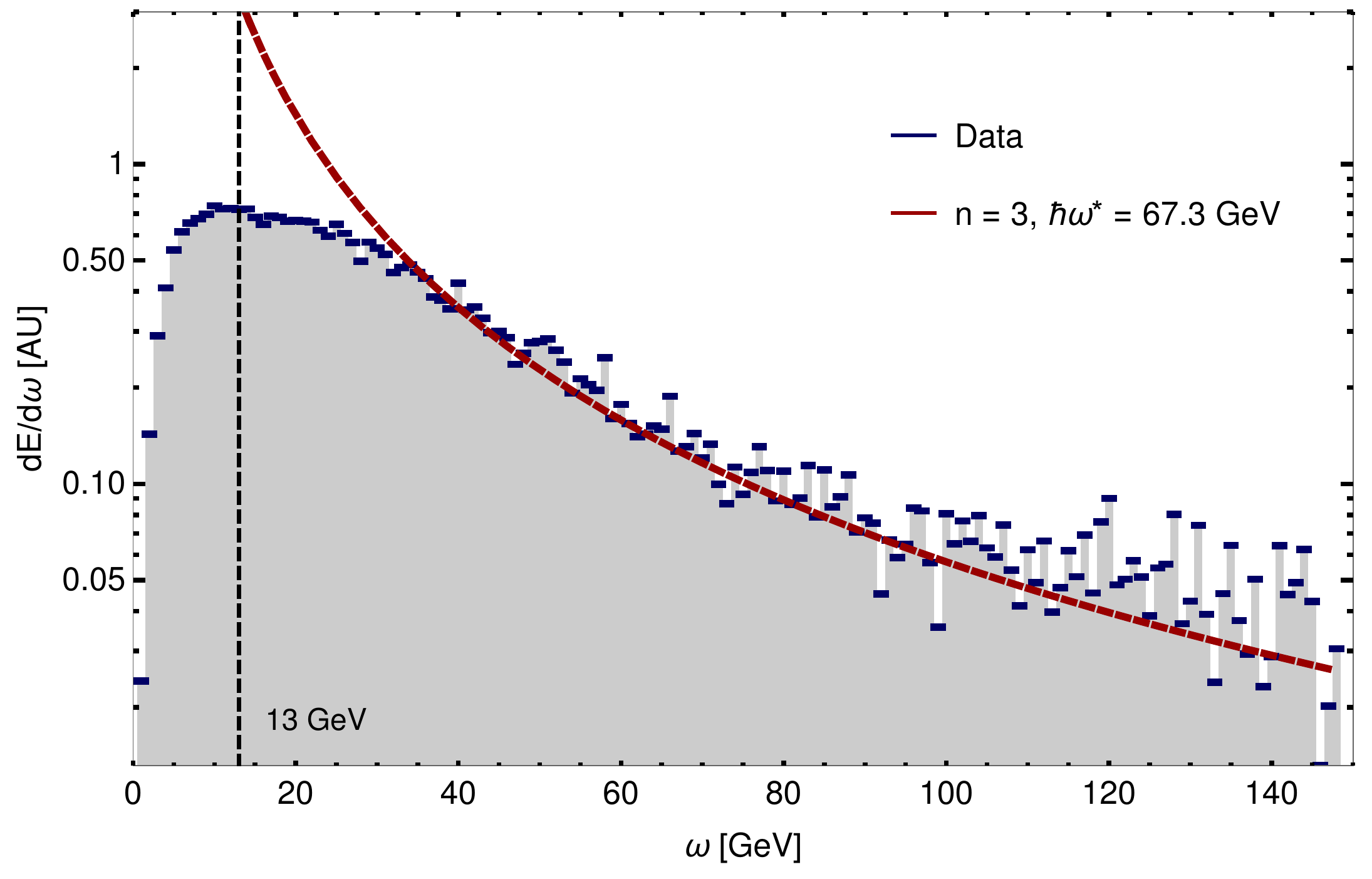}
\caption{Here we present the black hole energy spectrum, Eqn. (\ref{spec}), for n = odd. We find the best fit evaporation time threshold, $\hbar \omega^{\ast}$ = 67.3 GeV, and a chi-squared per degree of freedom $\chi^{2}_{3}/\nu = 1.88$.} 	
\label{plotoddbh}
\end{figure}

\begin{figure}[H]
\centering  
\includegraphics[scale=.23]{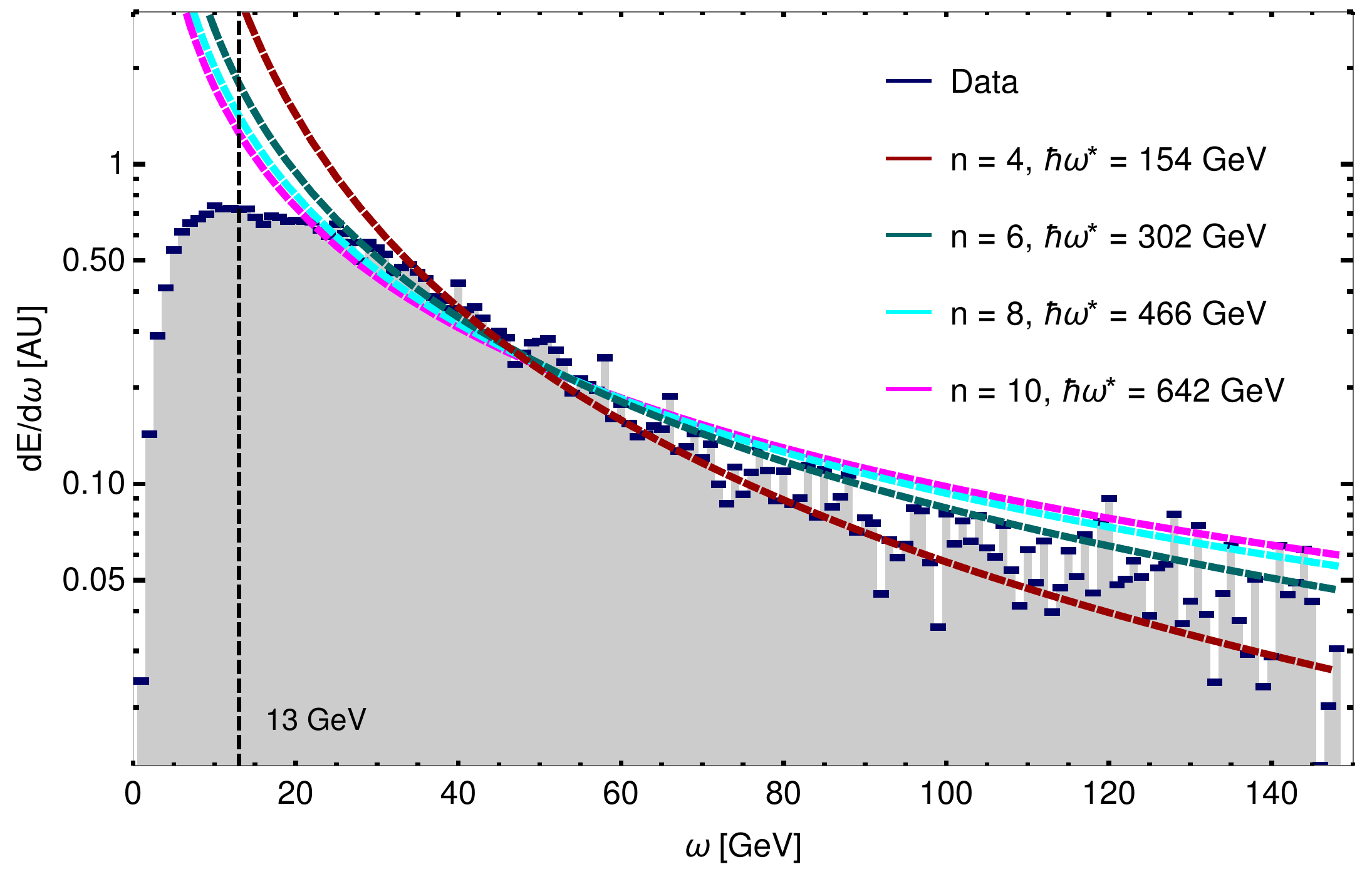}
\caption{Here we present the black hole energy spectrum, Eqn. (\ref{spec}), for n = even. We find the best fit evaporation time threshold, $\hbar \omega^{\ast}$, we have the following chi-squared per degree of freedom $\chi^{2}_{n}/\nu$: $\chi^{2}_{4}/\nu = 1.81$, $\chi^{2}_{6}/\nu = 2.28$, $\chi^{2}_{8}/\nu = 3.96$, and $\chi^{2}_{10}/\nu = 5.11$.} 	
\label{plotevenbh}
\end{figure}

\begin{figure}[H]
\centering  
\includegraphics[scale=.23]{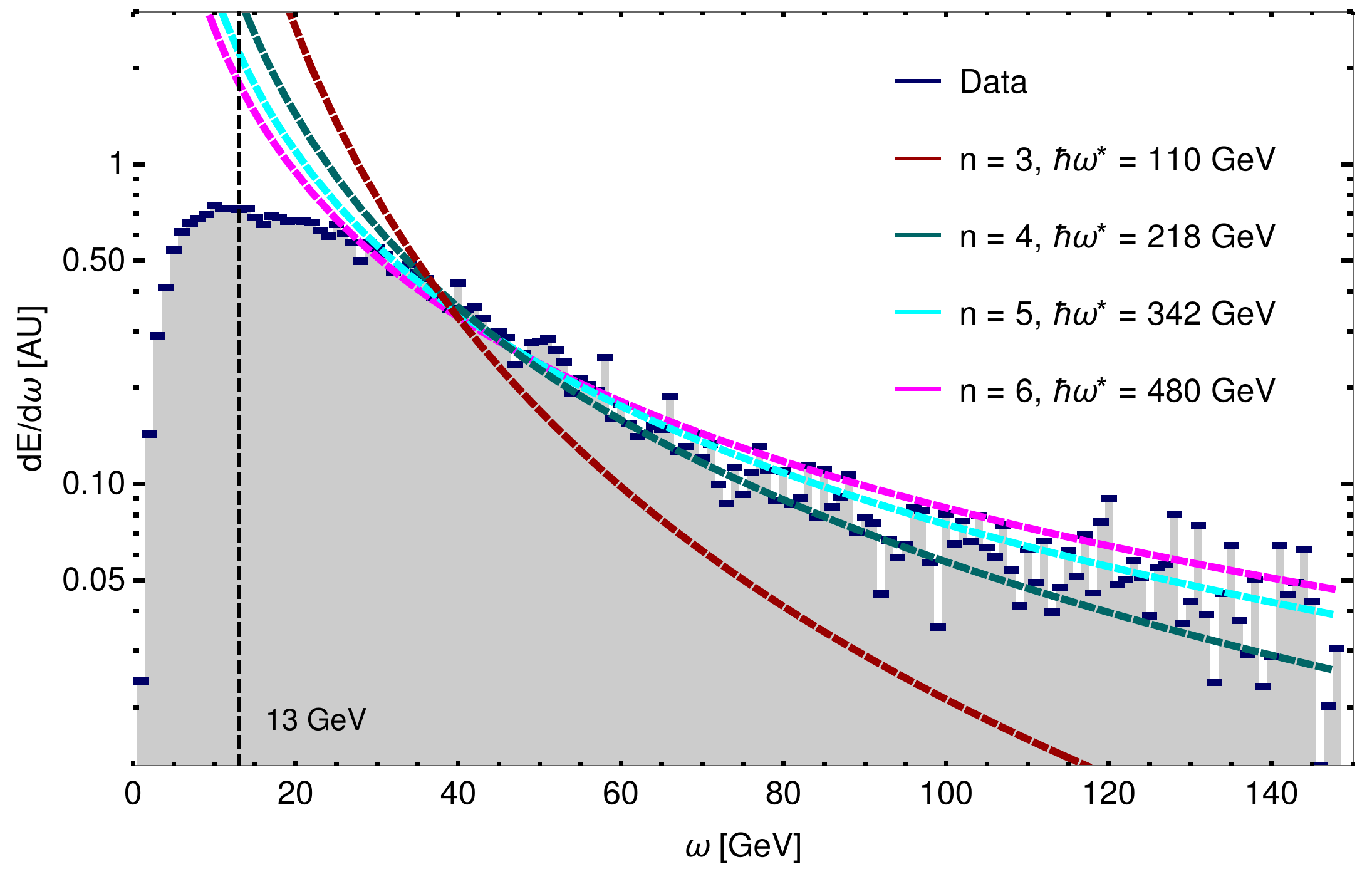}
\caption{Here we present the base black hole energy spectrum, Eqn. (\ref{spec0}), for $3 \leq n \leq 6$. We find the best fit evaporation time threshold, $\hbar \omega^{\ast}$, we have the following chi-squared per degree of freedom $\chi^{2}_{n}/\nu$: $\chi^{2}_{3}/\nu = 12.9$, $\chi^{2}_{4}/\nu =1.88$, $\chi^{2}_{5}/\nu = 1.44$, $\chi^{2}_{6}/\nu = 2.30$.} 	
\label{plotbarebh}
\end{figure}

There has been much interest in the dimensionality of black holes as well as the possibility of creating them at CERN \cite{PhysRevLett.87.161602}. In this regard, we shall examine the power radiated by the trans-Plankian diphoton black holes which, we hypothesize, comprise the thermal bath associated with the Unruh effect. For $n$ spatial dimensions, the $n$-dimensional Stefan-Boltzmann power law for a radiating hypersphere of radius, $R$, is given by \cite{landsberg1989stefan},
\bqe
\mathcal{P}_{n} = \frac{2n\zeta{(n+1)}}{\pi}\lb \frac{k_{B}}{\hbar c} \rb^{n} ck_{B} R^{n-1}T^{n+1}.
\eqe
Here, $\zeta(x)$ is the standard Riemann zeta function. 
Now, recalling the radius of the black hole is given by $R = \frac{n-2}{4 \pi T}\lb  \frac{\hbar c}{k_{B}} \rb$, we then note the product $(RT_{H})^{n-1} = \lb \frac{n-2}{4 \pi } \lb \frac{\hbar c}{k_{B}} \rb \rb^{n-1}$. Combining everything together yields a power law which, for any dimesions, radiates as $\sim T_{H}^2$. Hence,
\bqe
\mathcal{P}_{n} = \frac{2n}{\pi} \zeta{(n+1)}\lbk  \frac{n-2}{4 \pi} \rbk^{n-1}  \frac{k_{B}^2}{\hbar } T_{H}^{2}.\label{power}
\eqe
What is interesting to note, is that since this power law radiates as $T_{H}^{2}$, then the photon frequency dependence will go as $ \sim \omega^{-\frac{2}{n-2}}$. As such, the dimensionality of the black holes will be encoded in the steepness of the spectrum. The above expression is the power that is radiated by each diphoton black hole in the Rindler bath. The total evaporation time, $\Delta \tau$, for each black hole can be obtained by direct integration,
\bqe
\Delta \tau = \frac{(n-2)}{n}\frac{E_{R}}{\mathcal{P}_{n}}.\label{timeb}
\eqe
We also expect that there should be a thermal distribution, at the FDU temperature, of such black holes. However, just as the dimension of the black hole dictates the spectral steepness, so too does the dimensionality of the Rindler bath dictate the thermal statistics \cite{1986PThPS..88....1T}. For even spatial dimensions we will have Fermi-Dirac statistics and for odd spatial dimensions we will have Bose-Einstein statistics. Taking into account detailed balance we will have both spontaneous emission and stimulated decay. Summing over both will yield the following thermal factors \cite{Cozzella:2017ckb, lynch2021experimental},
\bqa
\lbk \frac{1}{e^{\frac{E_{R}}{k_{B}T_{FDU}}} + (-1)^n }\rbk  &+&  \lbk\frac{1}{e^{\frac{E_{R}}{k_{B}T_{FDU}}}+ (-1)^n }+1 \rbk  \non \\ &\approx & \frac{2k_{B}T_{FDU}}{E_{R}}, \;\; n=odd\non \\
&\approx & 2, \;\;\;\;\;\;\;\;\;\;\;\;\;\;\;\; n=even.
\eqa
Finally, we must weight our power radiated by the appropriate dimensional thermal factor. We must also boost the FDU temperature back to the lab frame \cite{lynch2023experimental}, $T_{FDU}\rightarrow T_{FDU}/\gamma$. Hence,
\bqa
\mathcal{P}_{odd} &=& \frac{4n}{\pi} \zeta{(n+1)}\lbk  \frac{n-2}{4 \pi} \rbk^{n-1}  \frac{k_{B}^3}{\hbar }  \frac{T_{H}^{2}T_{FDU}}{\gamma E_{R}} \non \\
\mathcal{P}_{even} &=& \frac{4n}{\pi} \zeta{(n+1)}\lbk  \frac{n-2}{4 \pi} \rbk^{n-1}  \frac{k_{B}^2}{\hbar } T_{H}^{2}.
\eqa
We must now formulate the power spectrum, $\frac{d\mathcal{P}_{n} }{d(\hbar \omega)}$. This can be accomplished by noting the following derivatives; $\frac{d T_{H}}{d(\hbar \omega)} = -\frac{T_{H}}{n-2} \frac{1}{\hbar \omega}$, $\frac{dT_{FDU}}{d(\hbar \omega)} =  \frac{ 2T_{FDU}}{\hbar \omega}$, and $\frac{d E_{R}}{d(\hbar \omega)} =\frac{E_{R}}{\hbar \omega}$. With these derivatives, we then have $\frac{d}{d(\hbar \omega)} \lb \frac{T_{H}^{2}T_{FDU}}{E_{R}}  \rb = -\frac{T_{H}^{2}T_{FDU}}{(\hbar \omega)E_{R}} \lb \frac{4-n}{n-2}  \rb$ and $\frac{d T_{H}^2}{d(\hbar \omega)} = -\frac{2T_{H}^2}{(\hbar \omega)(n-2)} $. Thus, combining all the appropriate pieces and weighting our power spectrum by the appropriate thermal distribution factors, our power spectrum will then take the following form,
\bqa
\frac{d\mathcal{P}_{odd} }{d(\hbar \omega)} &=& \frac{n(4-n)}{\pi^2} \zeta{(n+1)}\lbk  \frac{n-2}{4 \pi} \rbk^{n-2}  \frac{k_{B}^3}{\hbar } \frac{T_{H}^{2}T_{FDU}}{\gamma (\hbar \omega)E_{R}} \non \\
\frac{d\mathcal{P}_{even} }{d(\hbar \omega)} &=& \frac{2n}{\pi^2} \zeta{(n+1)}\lbk  \frac{n-2}{4 \pi} \rbk^{n-2}  \frac{k_{B}^2}{\hbar } \frac{T_{H}^2}{(\hbar \omega)}.
\eqa
Here, we have dropped the overall minus signs since the power radiated by the black hole is minus the power radiated by the positron. What is interesting to note is that for odd spatial dimensions, we are restricted to $n = 3$ in order to have a positive definite power spectrum. This prescription is by necessity in that had we not defined the black hole power as being the negative of the power radiated by the positron, then the dimensionality, for both odd and even dimensions, would have strictly ruled out the $n=3$ case. This situation seems rather unphysical given that we know 3 dimensional black holes exist \cite{PhysRevD.108.044047} and their existence does not depend on their statistical distribution. In order to compare the above power spectra, to the NA63 data set we must multiply by the average black hole evaporation time Eq.~(\ref{timeb}), $\Delta \overline{\tau} $, to produce the energy spectra,
 \bqa
\frac{dE_{odd} }{d(\hbar \omega)} &=& \frac{n(4-n)}{\pi^2} \zeta{(n+1)}\lbk  \frac{n-2}{4 \pi} \rbk^{n-2}  \frac{k_{B}^3}{\hbar } \frac{T_{H}^{2}T_{FDU}\Delta \overline{\tau}}{\gamma (\hbar \omega)E_{R}} \non \\
\frac{dE_{even} }{d(\hbar \omega)} &=& \frac{2n}{\pi^2} \zeta{(n+1)}\lbk  \frac{n-2}{4 \pi} \rbk^{n-2}  \frac{k_{B}^2}{\hbar } \frac{T_{H}^2\Delta \overline{\tau}}{(\hbar \omega)}.\label{spec}
\eqa
We must also note that if the properties of the radiation emitted by these diphoton black holes does not depend on the statistics of the Rindler bath, then our base power spectrum, $\frac{d\mathcal{P}_{0}}{d(\hbar \omega)}$, will be the same as the case for n = even up to an overall scaling factor of 2. Thus,
\bqe
\frac{dE_{0} }{d(\hbar \omega)} = \frac{n}{\pi^2} \zeta{(n+1)}\lbk  \frac{n-2}{4 \pi} \rbk^{n-2}  \frac{k_{B}^2}{\hbar } \frac{T_{H}^2\Delta \overline{\tau}}{(\hbar \omega)}.\label{spec0}
\eqe
These spectra, for various spatial dimensions, $n$, are presented in Fig. (\ref{plotoddbh}), (\ref{plotevenbh}), and (\ref{plotbarebh}) and compared to the data set. For all cases we have used $\beta_{\perp} = .012$ \cite{lynch2019accelerated} and $\alpha_{T} = \frac{1}{\pi}$ \cite{lynch2023experimental}. It is clear that $n=3$ for the odd case and $n=4$ for both the even n and base case is favored. This is due to the fact that for these cases, the spectrum will go as $\sim \omega^{-2}$. 

Let us examine, in more detail, the relevant chi-squared statistics and evaporation time threshold, $\hbar\omega^{\ast}$, for each of our black hole power spectra. Recalling the Unruh radiation analysis \cite{lynch2021experimental} that our chi-squared statistic goes to 1 between the regions 30 - 120 GeV. The low energy cutoff at 30 GeV is due to the thermalization time threshold, computed from the best fit power spectrum, of about 22 GeV. This small transient regime, between 22 and 30 GeV, reflects the onset of thermalization. The deviation of the chi squared at about 120 GeV is, in all likelihood, due to higher order processes and the subject of ongoing investigation. In this regard, we compute our chi-squared statistics for the black hole power spectra in the same 30 - 120 GeV window. Next, in examining the evaporation time threshold, we must recall that in both the Unruh radiation analysis \cite{lynch2021experimental} as well as the 1-d Planck moving mirror observation \cite{10.1093/ptep/ptad157}, the average acceleration which sets the temperature scale of both of these systems. As such, we should expect an average evaporation time threshold, $\hbar \omega^{\ast}$, to determine the overall black hole evaporation time, $\Delta \overline{\tau}(\hbar \omega^{\ast})$. This threshold is found via best fit for each black hole power spectrum. We can also compute the average photon frequency of our data set and we expect this average to characterize the data in the same way the average acceleration sets the radiation analysis. As such, from the experimental bin values, we have a thermalization threshold of 13.1 GeV and an upper bound on the data of 149 GeV. The computation of the average and variance yields $\hbar \omega^{\ast} = 68.2 \; \pm 38.3$ GeV. Thus, we expect our threshold scale to reside in this window. We do indeed find that the standard $n_{odd} = 3$ case does indeed fit this criterion. A table of the best chi-squared fits for our black hole power spectra, for various dimensions, and their associated threshold energy is presented in Table~\ref{chit}.
\begin{table}[ht] 
\centering
\begin{tabular}{l | l | l | l | l}
\hline\hline
statistics & \;odd\; & \;even\; & \;base\; & \;base\;\\ \hline
\;\;\;\;\;\;n & \;\;\;3  &  \;\;\;4  & \;\;\;4  & \;\;\;5 \\ \hline

\;\;\;$\chi^2/\nu$ 		&\;1.88\; &\;1.81\; &\;1.88\; &\;1.44\; \\ 
$\hbar \omega^{\ast}$ [GeV] 	& \;67.3\; & \;154\; & \;218\; & \;342\; \\
\end{tabular} \\ 
\caption{The reduced $\chi^2$ per degree of freedom for the more accurate best fits and their associated evaporation time threshold, $\hbar \omega^{\ast}$. Note, the evaporation time threshold is set by the average photon frequency set by the bandwidth beyond the thermalization threshold. From the mean and variance of the experimental data, we $\hbar\omega^{\ast} = 68.2 \; \pm \; 38.3$ GeV. As such,with all other thresholds outside the experimental bounds on the photon frequency, we find the $n_{odd} = 3$ to be a viable candidate.} 
\label{chit}
\end{table} 
\begin{figure}[H]
\centering  
\includegraphics[scale=.23]{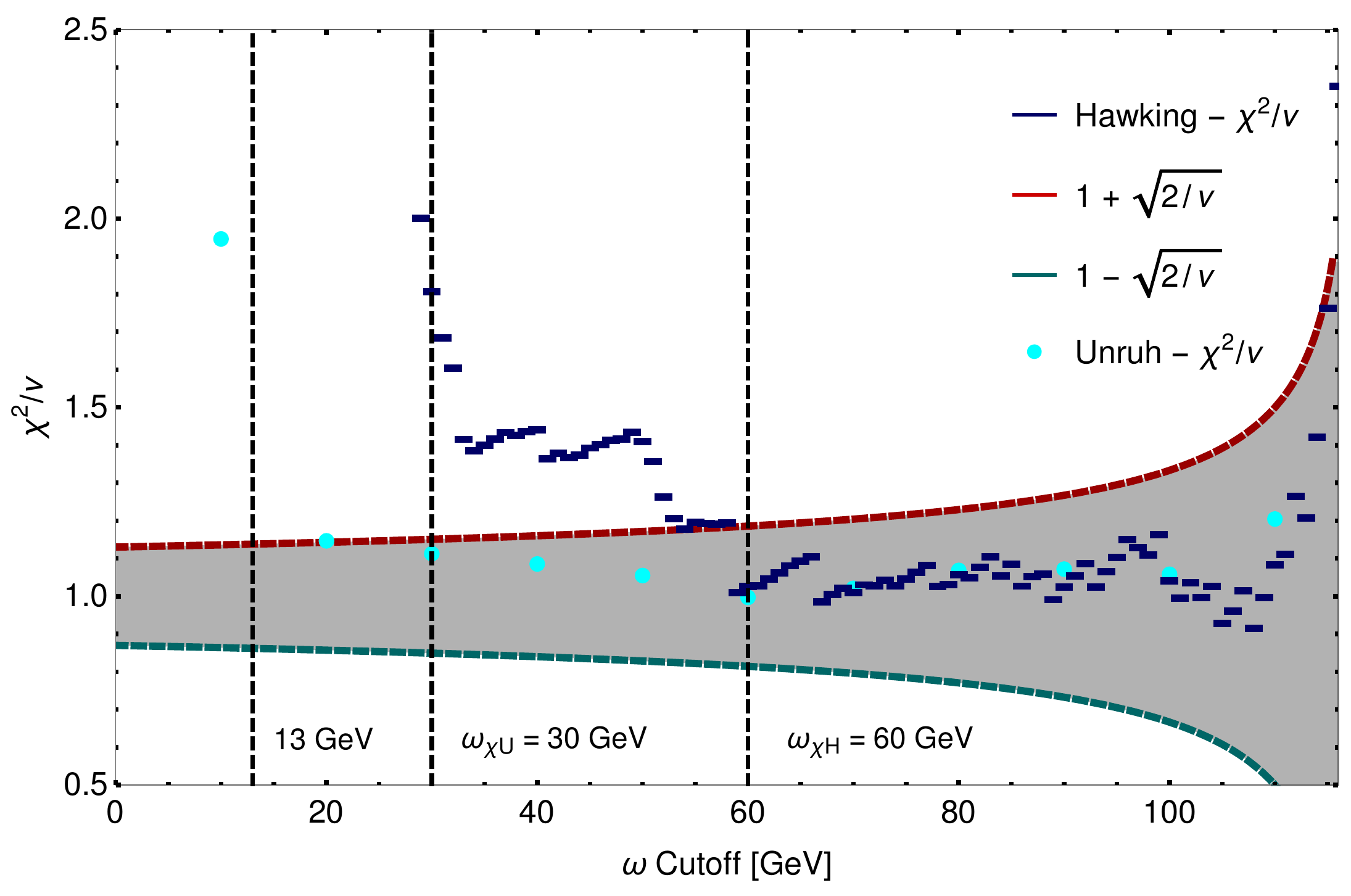}
\caption{The reduced chi-squared per degree of freedom of the standard $n_{odd} = 3$ Stefan Boltzmann energy spectrum compared to the Unruh effect radiation analysis \cite{lynch2021experimental}. Here, $13$ GeV is the thermalization time threshold, $\hbar \omega_{\chi U} = 30$ GeV is the chi-squared threshold of the Unruh effect analysis. Note the BH case goes below the 1 standard deviation threshold at $\sim 2 \hbar \omega_{\chi U} = \hbar \omega_{\chi H}=60$ GeV, i.e. at the associated diphoton energy.} 	
\label{plot6}
\end{figure}
The reduced chi-squared per degree of freedom as a function of low energy cutoff for the $n_{odd} = 3$ case is presented in Fig. (\ref{plot6}). What is interesting to note is that in the radiation analysis of the Unruh effect \cite{lynch2021experimental}, the chi-squared converges to the 1-standard deviation threshold at $\hbar \omega_{\chi} = 30$ GeV. Then, the Hawking analysis also converges to the 1-standard deviation threshold at $\sim 2\hbar \omega_{\chi}$, which further corroborates the presence of the diphoton. In both cases, the chi-squared statistic deviates outside the threshold above $\sim 120$ GeV. 

Let us finalize this section by analyzing the relevant length scales of the system. The overarching length scale which defines this system is set by the proper acceleration, $x_{a} = c^2/a$. For the average photon frequency of $\hbar\omega^{\ast} = 68.2$ GeV, we then have an acceleration of 89 PeV, and thus a distance scale of $x_{a} = 2.2 \times 10^{-24}$ m. As such, it appears that the down to these length/energy scales, the trans-Planckian black holes which, under assumption here, comprise the Rindler bath are described by the standard 3+1 dimensional spacetime. This is consistent with high energy cosmic rays \cite{2002PhRvD..65l4027A} and, in fact, pushes back the energy scale of extra dimensions to $M_{D} > 89$ PeV. Although we have not found convincing evidence for extra dimensions, these systems do raise the intriguing possibility of exploring various aspects of gravitation in particle physics. As an example, we have found evidence for a trans-Planckian black hole Rindler bath, as such we can then utilize this description to measure Newtons constant directly from the data set.

\section{Measurement of Newtons constant}
\begin{figure}[H]
\centering  
\includegraphics[scale=.23]{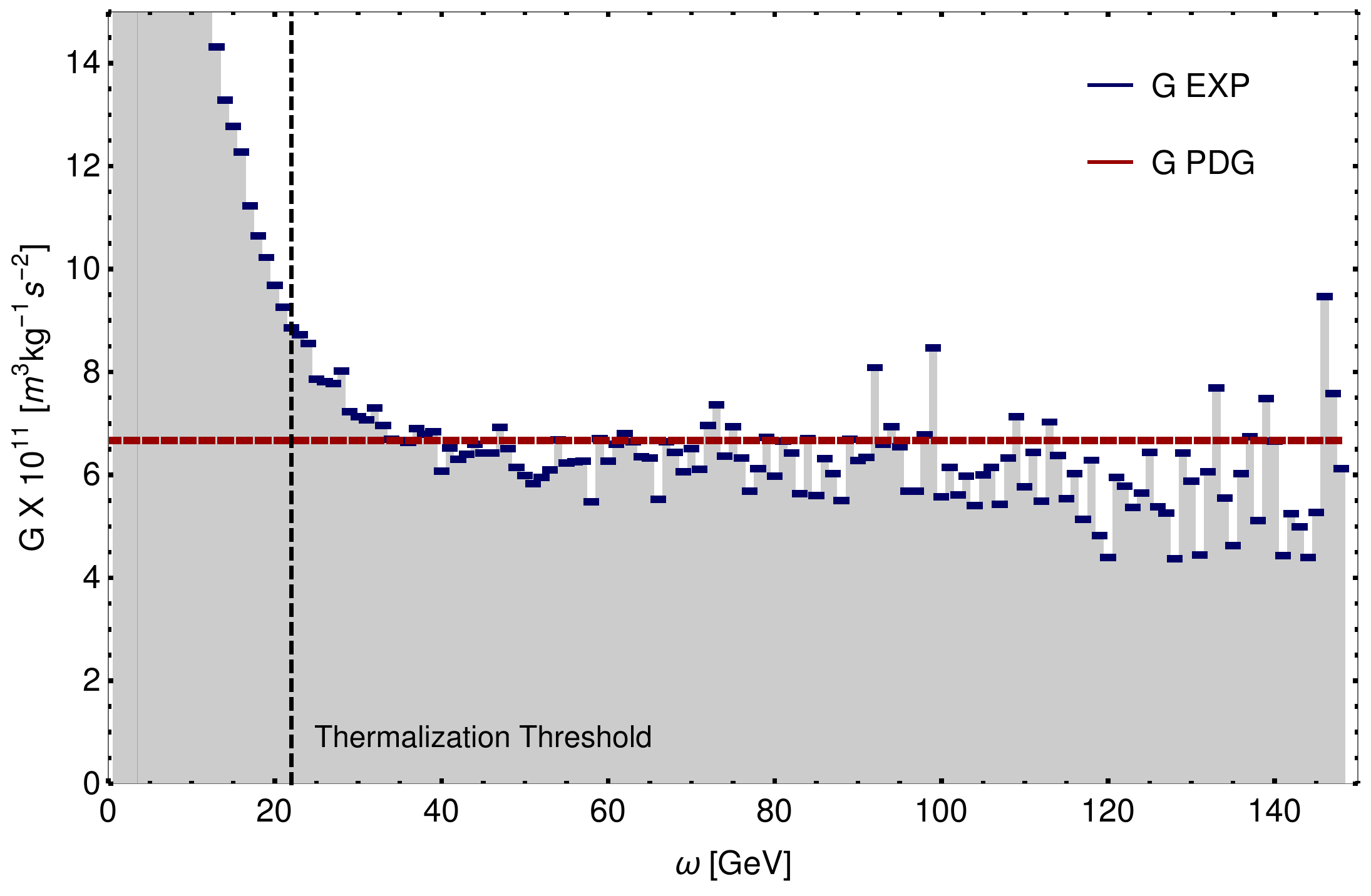}
\caption{Here we present the direct measurement of Newton's constant, Eqn. (\ref{bigg}), from the NA63 data set. We find $G=6.33 \pm .616 \times 10^{-11}$ $m^{3}kg^{-1}s^{-2}$. As such, we confirm the presence of gravitation in high energy channeling/radiation reaction experiments.} 	
\label{plot7}
\end{figure}

Given the fact that our most viable candidate, the $n_{odd} = 3$ spectrum, Eqn. (\ref{spec}), depends on the Hawking temperature, we can then utilize the data set to provide a direct measurement of Newton's constant, $G$. In doing so, we confirm the presence of gravitation in these systems and thus provide a novel experimental platform for probing gravity. In particular, the high statistics typically associated with particle physics may indeed yield high precision measurements of Newtons constant, thereby shedding light on the problem of its precision measurement \cite{10.1093/nsr/nwaa165,2007Sci...315...74F}. The Solving this spectrum for $G$ yields,
\bqe
G = \sqrt{\frac{1}{122880} \frac{c^8}{ \alpha_{T}\beta_{\perp}^3 \pi^2}\frac{\hbar}{m}\frac{\Delta t}{\gamma}\frac{1}{\frac{dE}{d(\hbar \omega)}}}. \label{bigg}
\eqe
The measurement of Newton's constant is presented in Fig. (\ref{plot7}). Here we have used $\alpha_{T} = \frac{1}{\pi}$, $\beta_{\perp} = .012$, and the best fit evaporation time threshold, $\hbar \omega^{\ast} = 67.3$ GeV. From the data set, we average the measured value of $G $ over the region, $30-120$ GeV. This is where the chi-squared statistic remains below the 1-standard deviation threshold in the radiation analysis \cite{lynch2021experimental}. Over this region, we obtain a value of $G=6.33 \pm .616 \times 10^{-11}$ $m^{3}kg^{-1}s^{-2}$. The accepted value of the Newton's constant from the particle data group (PDG) is $G = 6.67 \times 10^{-11}$ $m^{3}kg^{-1}s^{-2}$ \cite{Workman:2022ynf}.

\section{conclusions}
In this manuscript we examined the Unruh effect, measured by the high energy channeling experiments of CERN-NA63, from the point of view of the diphoton quantum fluctuations which comprise the Rindler bath. Under the assumption that these diphoton pairs are born out of the Hawking evaporation of microscopic trans-Planckian black holes, we find the resultant heat capacity and $n$-dimensional Stefan-Boltzmann energy spectrum in excellent agreement with the data. The resultant chi-squared analysis also places the theory within the 1 standard deviation threshold. In particular, we find a thermalization energy scale of the Hawking radiation at twice that of the Unruh radiation, thereby demonstrating consistency across mutually exclusive analyses. Moreover, the dimensionality of the energy spectrum demonstrates that the diphoton black hole, and subsequent Hawking radiation, is consistent with a 3+1 dimensional spacetime down to $\sim 89$ PeV. We finalize the analysis with a direct measurement of Newton's constant, and thereby provide confirmation of the presence of gravitation at NA63.

\section*{Acknowledgments}
This work has been supported by the National Research Foundation of Korea under Grants No.~2017R1A2A2A05001422 and No.~2020R1A2C2008103.


\bibliography{ref}

\end{document}